\documentclass[prl,twocolumn,showpacs,preprintnumbers,amsmath,amssymb]{revtex4}

\usepackage{amsmath,amssymb,amsfonts} 	
\usepackage[dvips]{graphicx}
\usepackage[latin1]{inputenc}
\usepackage{subfigure}
\begin{document}

\title{Origin of Power Laws for Reactions at Metal
  Surfaces Mediated by Hot Electrons}

\author{Thomas Olsen}
\author{Jakob Schi\o tz}
\email{schiotz@fysik.dtu.dk}

\affiliation{Danish National Research Foundation's Center of Individual
Nanoparticle Functionality (CINF),
	Department of Physics,
	Technical University of Denmark,
	DK--2800 Kongens Lyngby,
	Denmark}

\date{\today}

\begin{abstract}
  A wide range of experiments have established that certain chemical
  reactions at metal surfaces can be driven by multiple hot electron mediated 
  excitations of adsorbates. A high transient density of hot
  electrons is obtained by means of femtosecond laser pulses and a
  characteristic feature of such experiments is the emergence of a
  power law dependence of the reaction yield on the laser fluence
  $Y\sim F^n$. We propose a model of multiple inelastic scattering by
  hot electrons, which reproduces this power law and the experimentally
  found exponents of several experiments. All parameters are
  calculated within Density Functional Theory and the Delta
  Self-Consistent Field method. With a simplified assumption, the power law becomes
  exact and we obtain a simple and very useful physical interpretation
  of the exponent $n$, which represents the number of
  adsorbate vibrational states participating in the reaction.
\end{abstract}

\pacs{82.53.St, 34.35.+a}
\maketitle
Hot-electron induced femtochemistry at surfaces (HEFatS) is a class of
chemical reactions, where the energy to overcome the reaction barrier
is provided by energetic ("hot") electrons.  These electrons are often
photoelectrons emitted from the surface when struck by intense laser
pulses, but other possible hot-electron sources are
Scanning Tunneling Microscope tips and Metal-Insulator-Metal
nanodevices \cite{gadzuk96}. For these reactions, the reaction rate usually
scales as a power law with the electron flux. For example, in a pioneering
study of NO on Pd(111) \cite{prybyla1_etal}, it was shown that femtosecond
laser pulses could induce desorption of NO, and a mechanism by
multiple electronic excitations was identified \cite{budde_etal}.  The
yield depended on the fluence as $Y\propto F^n$ with
$n\sim3.3$. Subsequently, desorption induced by femtosecond lasers has
been demonstrated for several other adsorbate systems
\cite{prybyla2,kao2,struck_etal,ho} showing non-linear yields, which can be
fitted to power laws with exponents $3<n<8$. It has also been shown
that femtosecond laser pulses can induce surface hopping \cite{stepan}
and oxidation reactions \cite{bonn1,kao1,deliwala_etal} all of which can be
characterized by power laws.

A popular theoretical approach to the interaction of adsorbates with a high density of hot electrons is the electronic friction model \cite{brandbyge_etal} where the excited electrons are assumed to equilibrate rapidly to a thermal distribution with electronic temperature $T_e$. For sufficiently large timescales the propagation of the adsorbate is well approximated by semiclassical Langevin dynamics with an electronic friction $\eta_e$ depending on $T_e$ as well as the adsorbate coordinates. While the friction model has certainly been successful in reproducing various experimental observations, it cannot account for the physical origin of the power law exponent $n$. Furthermore, measurements on the distribution of hot electrons in gold excited by subpicosecond laser pulses imply thermalization times up to $1\;ps$ \cite{fann1_etal}, which is on the order of reaction times, and there are examples of laser induced surface chemistry where the assumption of a thermal distribution of hot electrons is in direct conflict with observations \cite{kao2,busch,deliwala_etal}.

In this letter we introduce a general model for first-principles
calculations of the rates of HEFatS processes, regardless of the
source of hot electrons.  The central object of the model is a
non-adiabatic Hamiltonian, which is used to calculate the hot electron
induced vibrational transition probabilities of adsorbates. All
parameters in the Hamiltonian are obtained from density functional
theory and we show that the model reproduces experimentally observed
power laws. Finally we will make a simple approximation for the
transition probabilities and show that the power law then becomes
exact in the limit of large electron flux and that the exponent is
given by $n=E_R/\hbar\omega$ where $E_R$ is the energy barrier of the
reaction and $\hbar\omega$ is the energy quantum of the vibrational
mode dominating the energy transfer.

To analyze HEFatS rates involving multiple hot electrons, we consider a quadratic potential energy surface of an adsorbate with vibrational states $|n\rangle$ coupled to a localized electronic resonance $|a\rangle$. The probability for a hot electron with energy $\varepsilon$ to inelastically scatter on the localized state and induce a vibrational transition $m\rightarrow n$ is denoted $P_{mn}(\varepsilon)$ and we assume a constant flux $J_0$ of hot electrons incident on the adsorbate. It is further assumed that each vibrational quantum has a fixed lifetime $T_{vib}$ and that there exist a maximum quantum number $n_R$ such that a reaction will proceed immediately if $n\geq n_R$. The probability that one vibrational quantum survives the time interval $\Delta t=1/J_0$ between subsequent scattering events is then $e^{-\Delta t/T_{vib}}$. Each incoming electron will thus change the distribution of adsorbate vibrational states $Q(n)$ until an equilibrium is reached between decay and reexcitation. 

If the adsorbate is initially in the ground state, the distribution of vibrational states induced by the first electron with energy $\varepsilon_1$ is
\begin{equation}\label{Q_1}
Q_1(n;\varepsilon_1)=P_{0n}(\varepsilon_1).
\end{equation}
The probability of the adsorbate being in the $n$'th vibrational state after the second electron has scattered is
\begin{equation}
Q_2(n;\varepsilon_1,\varepsilon_2)=\sum_{m=0}^{n_R-1}p_1(m;\varepsilon_1)P_{mn}(\varepsilon_2),
\end{equation}
where $p_1(m)$ is the probability that the adsorbate is in the state $m$ after the time interval $\Delta t$ given by
\begin{align}\label{bin}
p_1(m;\varepsilon_1)=\sum_{l=m}^{n_R-1}&Q_1(l;\varepsilon_1)\binom{l}{m}(e^{-\Delta t/T_{vib}})^m\notag\\
&\times(1-e^{-\Delta t/T_{vib}})^{l-m}.
\end{align}
We exclude terms with $m\geq n_R$ since such excitations would have led to a reaction by assumption.
Proceeding like this, the probability $Q_3(n;\varepsilon_1,\varepsilon_2,\varepsilon_3)$ of being in the $n$'th excited state after the third scattering event can be expressed in terms of $Q_2(n;\varepsilon_1,\varepsilon_2)$ and so forth. Since all vibrational states $n\geq n_R$ lead to a reaction, the reaction probability of the $k$'th electron is
\begin{equation}\label{prob}
P^R_k=\sum_{n=n_R}^\infty Q_k(n).
\end{equation}
For large $k$ this will approach a limiting value, $P_R$.

To calculate the vibrational transition matrix $P_{mn}(\varepsilon)$ we consider a Newns-Anderson type Hamiltonian with substrate states $|k\rangle$ and a resonant state $|a\rangle$ linearly coupled to a number of vibrational modes with creation operators for vibrational quanta $b_i^\dag$ \cite{wingreen,gadzuk91}:
\begin{align}\label{H}
H&=\varepsilon_0c_a^{\dag}c_a + \sum_k\epsilon_kc_k^{\dag}c_k+\sum_k\Big(V_{ak}c_a^{\dag}c_k+V_{ak}^*c_k^{\dag}c_a\Big)\notag\\
&+\sum_i\hbar\omega_i(b_i^{\dag}b_i+\frac{1}{2})+\sum_i\lambda_ic_a^{\dag}c_a(b_i^{\dag}+b_i).
\end{align}
The model essentially describes a harmonic oscillator, which is displaced when the state $|a\rangle$ is occupied and the coupling $V_{ak}$ to the metallic states introduces a finite lifetime of $|a\rangle$. As previously published \cite{olsen1}, the transition probabilities $P_{mn}(\varepsilon)$ can be calculated exactly in the wide band limit where the density of states projected on the localized state $|a\rangle$ is a Lorentzian centered at $\varepsilon_0$ with full width at half maximum given by $\Gamma=2\pi\sum_k|V_{ak}|^2\delta(\varepsilon_0-\epsilon_k)$. The probabilities $P_{mn}(\varepsilon)$ depend on the dimensionless parameters: $\hbar\omega_i/\Gamma$ and $g_i=(\lambda_i/\hbar\omega)^2$ and the reaction probability $P^R$ also depends on these parameters in addition to the reaction quantum number $n_R\sim E_R/\hbar\omega$. The quantities $E_R$ and $\omega$ can be calculated within standard density functional theory and $\Gamma$ is estimated from the projected density of states. The resonance energies as well as the non-adiabatic coupling parameters $\lambda_i$ is obtained from excited state potential energy surfaces which are calculated with the method of linear expansion Delta Self-Consistent Field. The method is a generalization of standard Delta Self-Consistent Field designed to handle molecular orbitals hybridized with metallic states and calculates the expectation values of excited states which are not eigenstates of the Hamiltonian but involve an occupied resonance. Details on the method and comparison with experiment can be found in Refs. \cite{olsen1} and \cite{gavnholt}.

\begin{figure}[tb]
	\includegraphics[width=6.0 cm]{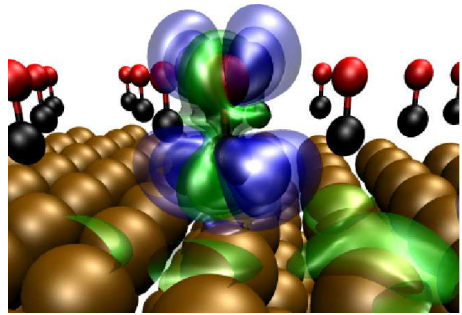} 
\caption{(Color) The difference between excited and ground state densities for CO adsorbed on Cu(111). Black balls are Carbon atoms and red balls are Oxygen atoms. Blue contours is excess density in excited state and green contours is excess density in ground state. The excited state is constructed by occupying a $\pi^*$ orbital of CO. For clarity we only show the density difference in a single supercell.}
\label{fig:density}
\end{figure}
\begin{figure}[b]
	\includegraphics[width=8.0 cm]{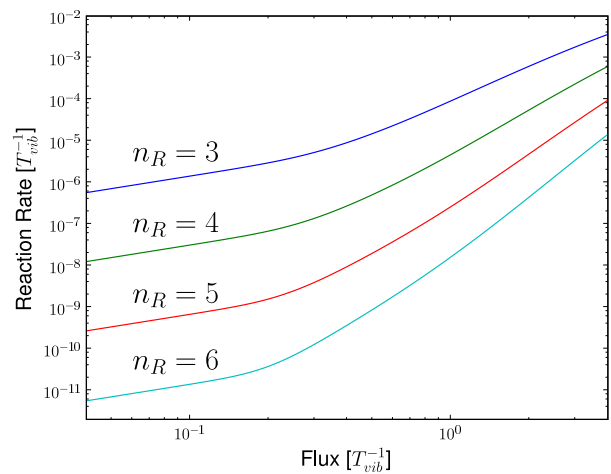} 
\caption{(Color online) The reaction yield of CO on Cu(111) as a function of hot electron flux for four reaction energies. Note the initial linear dependence corresponding to single electron reactions.}
\label{fig:CO}
\end{figure}        
As an example we calculate the fluence dependent transfer of energy
from hot electrons to a CO molecule adsorbed on Cu(111), mediated by
excitation of an unoccupied $\pi^*$ orbital. CO adsorbs with the
molecular axis perpendicular to the surface and the symmetry of the
adsorbed molecule thus only allows a linear coupling to the center of
mass (COM) and internal stretch vibrations of CO. The parameters for
CO chemisorbed at a Cu(111) bridge site were calculated with the code
\texttt{gpaw} \cite{gpaw,mortensen}, which is a real-space Density
Functional Theory code using the projector augmented wave method
\cite{blochl1}. We modelled the surface by a three layer ($4\times4$) supercell with the top layer relaxed, a grid-spacing of 0.2\AA, and a $4\times4$ surface K-point sampling. With the RPBE \cite{hammer} functional,
we find $\varepsilon_0=2.4\;eV$, $\hbar\omega=231\;meV$ and $\lambda=-118\;meV$ for the internal
stretch vibration, and $\hbar\omega=42\;meV$ and $\lambda=-4\;meV$ for
the COM vibration. 
The internal mode completely dominates the transfer of energy from hot
electrons to the molecule since it has a much larger coupling
$\lambda$, and the quantum of energy is five times larger than for the
COM mode.
Figure \ref{fig:density}
shows the density of an excited top site molecule obtained with
$\Delta$SCF-DFT relative to the ground state density and one clearly
sees the excited $\pi^*$ orbital. In figure \ref{fig:CO} we have used \eqref{Q_1}-\eqref{H}
with hot electrons at $\varepsilon=2.0\;eV$ corresponding to the laser
frequency used in \cite{prybyla2} to calculate reaction rates, which
require energies corresponding to 3, 4, 5 and 6 internal vibrational	
excitations. In the non-linear regime corresponding to reactions
induced by multiple scattering events, the rates are very well
approximated by power laws with $n=2.8$, $n=3.6$, $n=4.3$, and
$n=5.1$. Varying the parameters in the model reveals that in general,
one always obtains very good power law fits with exponents $n\sim
n_R$. Increasing the energy of the hot electrons above the resonant
energy $\varepsilon_0$ tend to decrease $n$.

\begin{table}[t]
\begin{center}
\begin{tabular}{c|c|c|c}
System & Experimental $n$ & Calculated $n$ & $E_R$\\
	\hline
NO/Pd(111) \cite{prybyla1_etal} & 3.3       & 3.7     & 1.0 \textit{eV}\\
CO/Cu(111) \cite{prybyla2}   & 3.7       & 3.6     & 0.4 \textit{eV}\\
CO/Cu(100) \cite{struck_etal} & $8.0\pm1$ & 3.7     & 0.5 \textit{eV}\\
O$_2$/Pt(111) \cite{kao2}  & 3.0/5.6   & 2.5/5.5 & 0.4 \textit{eV}\\
O/Pt(111) \cite{stepan}    & 15        & 12      & 0.8 \textit{eV}\\
\end{tabular}
\end{center}
\caption{Power law exponents obtained from \eqref{Q_1}-\eqref{H}.
  The systems involving NO, CO, and O$_2$ are
  desorption experiments whereas O/Pt(111) is hot electron induced
  diffusion of atomic oxygen from a step to a terrace hollow site.
  For O$_2$/Pt(111), the exponent depends on the laser
  frequency.}
\label{tab1}
\end{table}
Table \ref{tab1} shows the calculated power law exponents for five systems compared with corresponding experiments. The DFT parameters use were the same as for CO on Cu(111). We have assumed that a reaction occurs when an energy of $0.2\;eV$ in excess of the reaction barrier has been transferred, which is consistent with measurements of the kinetic and internal energy distributions of desorbed molecules \cite{struck_etal} and we have used hot electron energies $\varepsilon=\varepsilon_f+\hbar\Omega$ where $\Omega$ is the laser frequency. 
The agreement is very good except for CO on Cu(100). This could be due to the role of frustrated rotations, which has previously been found to couple strongly to metallic electrons in this particular system \cite{tully}. In general, however, the frustrated rotations have a low energy compared with the internal stretch mode and we expect their effect to be neglectable. Due to symmetry the frustrated rotations cannot couple linearly to the resonant electron and one would need a quadratic term like $\lambda_2c_ac_a^\dag(b+b^\dag)^2$ in \eqref{H} to include these in the model \cite{olsen2}.

A particularly interesting case is the hot electron induced desorption of O$_2$ from Pt(111) where a power law with exponent $n\sim5.6\pm0.7$ was observed using a photon energy of $2.0\;eV$ and an exponent of $n\sim3.0\pm0.5$ using a photon energy of $4.0\;eV$ \cite{kao2}. The fact that the power law exponent depends on the photon energy contradicts the picture of a thermalized hot electron gas interacting with the molecule, which is the basic assumption in models based on electronic friction. In contrast, the model 
\eqref{Q_1}-\eqref{H}
naturally gives rise to a decrease in the power law exponent when the energy of hot electrons is increased.

The transition matrix $P_{mn}(\varepsilon)$ has a very complicated structure and it is hard to extract the physics of the power law using these probabilities and \eqref{Q_1}-\eqref{prob}. However, the magnitude of $P_{mn}(\varepsilon)$ is largely governed by the prefactor $g^{n-m}/(n-m)!$ \cite{olsen1} and in the following we will examine the consequences of assuming transition probabilities of the form
\begin{equation}\label{P_mn}
P_{mn}=e^{-\alpha}\frac{\alpha^{n-m}}{(n-m)!},\qquad n\geq m,
\end{equation}
where $\alpha$ is a dimensionless coupling constant, which describes the coupling of hot electrons to the adsorbate vibrational states. Repeated use of the algorithm \eqref{Q_1}-\eqref{prob} with these probabilities then reveals that to leading order in $\alpha$ one has
\begin{align}\label{Q_n1}
Q_k(n)=\frac{\alpha^{n}}{n!}\bigg(\sum_{j=0}^{k-1}e^{-j\Delta t/T_{vib}}\bigg)^n.
\end{align}
We then consider a large flux $e^{-\Delta t/T_{vib}}\sim 1-\Delta t/T_{vib}$, sum up the geometric series, take the limit $k\rightarrow\infty$ corresponding to steady state, and get
\begin{align}\label{power}
Q(n)=\frac{\alpha^n}{n!}(T_{vib}J_0)^n,
\end{align}
where $\Delta t=1/J_0$. Thus, for small $\alpha$ the reaction probability \eqref{prob} will be dominated by such a term with $n=n_R$.

\begin{figure}[t]
          \includegraphics[width=4.25 cm, clip]{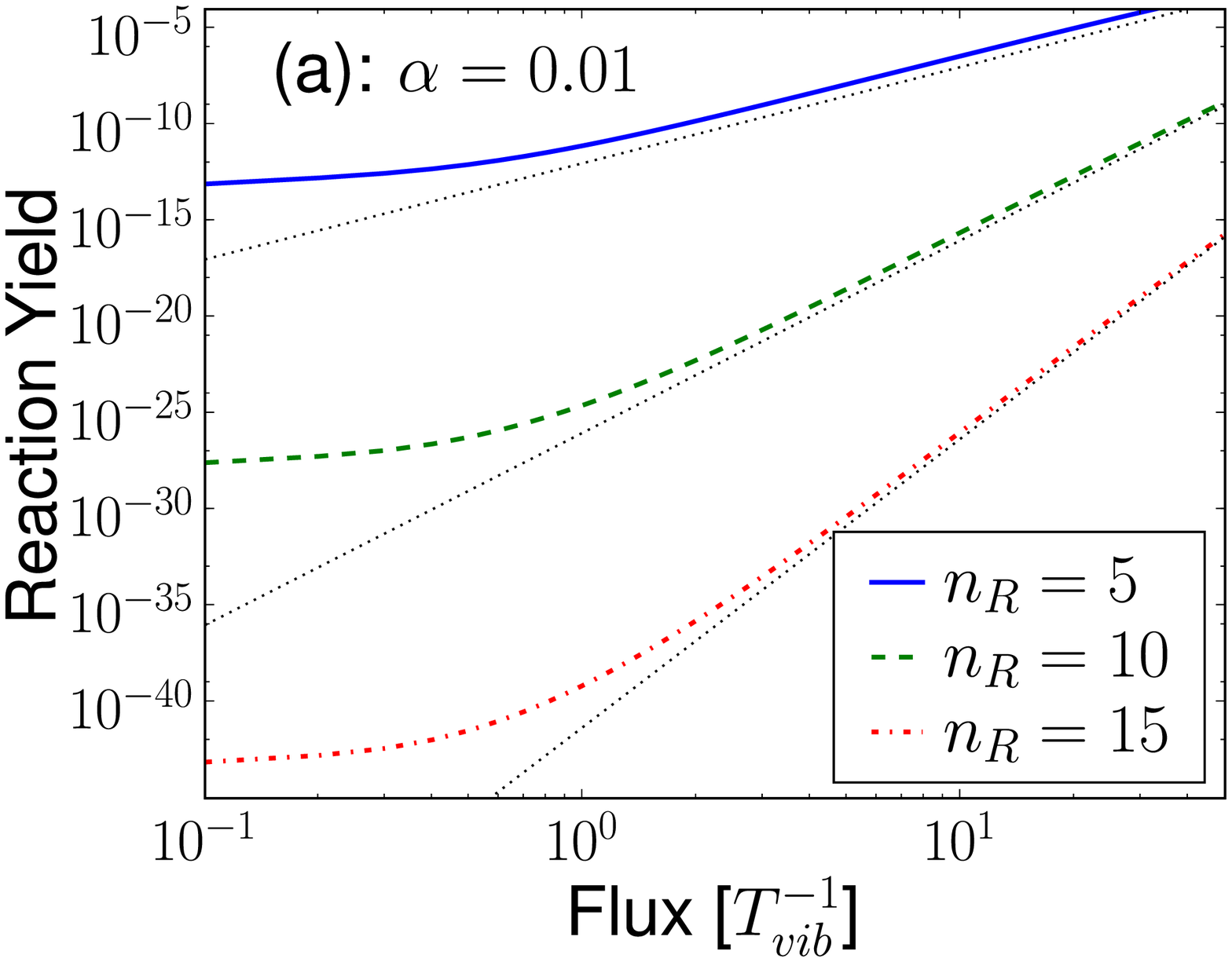}
          \includegraphics[width=4.25 cm, clip]{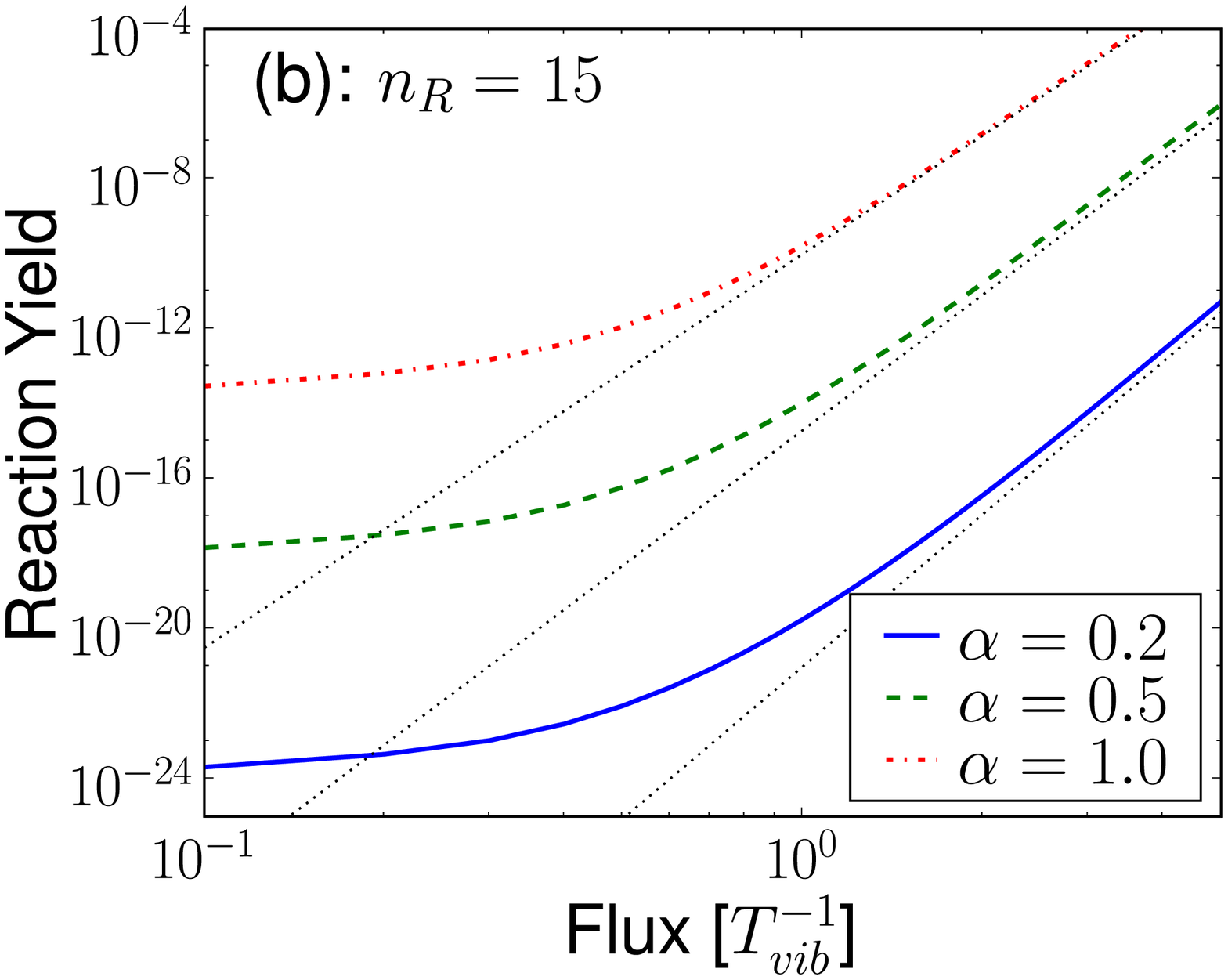}\\
\caption{(Color online) The yield as a function of electron flux obtained using the transition probabilities in equation \eqref{P_mn} for different parameters (see text).}
\label{fig:poisson}
\end{figure}
The power law emerges from summing up the detailed combinatorics of
all possible ways of rising through the vibrational states in the
potential well. 
In figure \ref{fig:poisson}a we show the reaction yields for three values of
$n_R$ and they are seen to approach power laws of the form $Y\propto
J_0^{n_R}$ for large fluxes. Even if $\alpha$ is not small, equations
\eqref{Q_1}-\eqref{prob} tend to conserve the power law although the
exponent becomes reduced from the value of $n_R$ when terms beyond
leading order are not vanishing. In figure \ref{fig:poisson}b we show
the yield when $n_R=15$ for $\alpha = 0.2$, $\alpha = 0.5$, and
$\alpha = 1.0$.  At large fluxes the yields are well approximated by
power laws with exponents 14, 12, and 10 respectively. One might worry
that the fixed time interval $\Delta t$ between scattering events is
too crude an approximation for the random nature of hot electrons
interacting with the adsorbate. However, a sequence of time intervals
$\{t_k\}$ with an average of $\Delta t$ would lead to the replacement
$j\Delta t\rightarrow \sum_{i=1}^jt_{k-i}$ in \eqref{Q_n1}, which is
well approximated by $j\Delta t$ for large $j$. 
We have repeated the calculations leading to figure \ref{fig:poisson}
but with the time intervals randomly drawn from an exponential distribution $p(t)\sim
e^{-t/\Delta t}$ and the power law is conserved on average.

The interpretation of the power law exponent as the number of contributing vibrational states can be used to identify the reaction channel of a given adsorbate system. For example, in the study of hot electron mediated desorption of NO from Pd(111) \cite{prybyla1_etal} a power law with exponent $n\simeq 3.3$ was found. The internal stretch vibration corresponds to an energy of $\hbar\omega\simeq210\;meV$ whereas the other modes have vibrational energies $\hbar\omega\leq70\;meV$. Since the adsorption energy is $E_a\simeq1.0\;eV$ we conclude that the power law exponent $n_R\sim E_{a}/\hbar\omega$ has to arise from sequential excitation of the internal stretch vibration and subsequent anharmonic energy transfer to the desorption coordinate. In contrast, the study of hot electron induced surface diffusion of atomic oxygen on Pt(111) \cite{stepan} gave rise to a power law with exponent $n\simeq15$, which fit very well with an experimental diffusion barrier of $E_d\simeq0.8\;eV$ and vibrational modes on the order $\hbar\omega\sim50\;meV$.

In summary, we have presented a theory of multiple inelastic
scattering with transition probabilities calculated in a non-adiabatic
Newns-Anderson model, which lead to the ubiquitous power law of HEFatS
and reproduce experimentally found exponents. The interpretation of
the exponents as the number of contributing vibrational states is a
very useful tool to identify the reaction channel of a given system
and also indicates that a classical treatmeant of the adsorbate motion
is insufficient. However, the model can only treat quadratic
potentials and are thus not able to treat the anharmonic effects, which
are expected to play an important role in the transfer of internal
vibrational energy to the reaction coordinate. In the case of molecular 
desorption the model applies because the internal stretch mode, which is well approximated by a harmonic potential,
dominates the energy transfer. Furthermore, while our assumption of constant hot electron energy
$\varepsilon=\varepsilon_f+\hbar\Omega$ gives the right dependence of
the exponents on $\Omega$, the hot electrons proceeding a laser pulse
will undergo scattering and produce some distribution of electron
energies. The true (time-dependent) distribution lies somewhere
between the present assumption and a thermalized hot electron gas and
the model should thus be regarded as complementary to a statistical
approach based on electronic friction and Langevin dynamics, which assumes a thermalized hot electron gas.

This work was supported by the Danish Center for Scientific Computing. The Center for Individual Nanoparticle Functionality (CINF) is sponsored by the Danish National Research Foundation.


\end{document}